\documentstyle[12pt,a4,bbbold,psfig]{article}

\def\dfrac#1#2{{\displaystyle#1\over\displaystyle#2}}
\def\slantfrac#1#2{\hbox{$\,^{#1}\!/_{#2}$}}

\begin{document}

\title{The impact of viscosity on the morphology of gaseous
flows in semidetached binary systems}
\author{
Bisikalo D.V.$^1$, Boyarchuk A.A.$^1$,\\
Kuznetsov O.A.$^2$, Chechetkin V.M.$^2$\\[0.3cm]
$^1$ {\it Institute of Astronomy of the Russian Acad. of Sci.,
Moscow}\\
{\sf bisikalo@inasan.rssi.ru; aboyar@inasan.rssi.ru}\\[0.3cm]
$^2$ {\it Keldysh Institute of Applied Mathematics, Moscow}\\
{\sf kuznecov@spp.keldysh.ru; chech@int.keldysh.ru}\\[0.3cm]
}
\date{}
\maketitle

\begin{abstract}

{\bf Abstract}---Results of 3D gas dynamical simulation of mass
transfer in binaries are presented for systems with various
values of viscosity.  Analysis of obtained solutions shows that
in the systems with low value of viscosity the flow structure is
qualitatively similar to one for systems with high viscosity
(see [1--6]). Presented calculations confirm that there is no
shock interaction between the stream from $L_1$ and the forming
accretion disk (`hot spot') at any value of viscosity.

\end{abstract}

\section*{INTRODUCTION}

Earlier we have already considered the morphology of
gaseous flows in semi\-deta\-ched binary systems [1--6]. Within
the framework of the 3D numerical simulations we
determined that the presence of rarefied circumbinary gas
substantially changes the structure of gas flows in the system.
In particular, the self-consistent solution does not include
shock interaction between the stream of gas from the inner
Lagrangian point $L_1$ and the forming accretion disk (`hot
spot'). However, these solutions were obtained for relatively
high viscosity in the disk because the insufficient power of the
computer used precluded fine grid calculations. In terms of the
$\alpha$-disk the numerical viscosity was
$\alpha\sim\slantfrac{1}{2}$. At the same time, the problem of
flow structure at low viscosity is of great interest. This is
primarily because the modern simulations of dwarf novae are
based on the assumption that the accretion disk has two states
with low and high viscosity (see, e.g., [7--10]). The
validity of this `limit cycle variation' model has been
confirmed by observational data (see, e.g., [10,11]).

The aim of this work is to investigate the general morphology
of gas flows in semidetached binary systems with low viscosity.
Particular attention will be paid to the problem whether the
flow structure remains the same with a decrease in viscosity,
and, in particular, whether the stream-disk interaction retains
its shock-free pattern as was shown in [1,2] for relatively
high viscosity. Study of stream-disk interaction and the related
problem of the absence of `hot spot' is of great importance for
the interpretation of the observations.

In present work we used Euler equations describing the flow of
unviscous gas (similar to works [1--6]). Nevertheless, the
behavior of the solution is influenced by the numerical
viscosity.  The presence of the only numerical viscosity and the
absence of the physical one in the simulation restricts detailed
investigation of the problem, while the dependence of the
solution on the viscosity can be described qualitatively.
Because numerical viscosity (assuming the finite--difference
scheme is already chosen) depends on the spatial and time
resolution, the viscosity dependence of the solution can be
obtained in sequential runs with decreasing size of the
gridcell.

\section{THE MODEL}

Let us consider the semidetached binary system with mass ratio
$q=1$. To describe the gas flow in this binary system we used
the 3D system of Euler equations

$$
\frac{\partial \rho}{\partial t}
+\frac{\partial \rho u}{\partial x}
+\frac{\partial \rho v}{\partial y}
+\frac{\partial \rho w}{\partial z}=0\,,
$$

$$
\frac{\partial \rho u}{\partial t}
+\frac{\partial (\rho u^2+P)}{\partial x}
+\frac{\partial \rho uv}{\partial y}
+\frac{\partial \rho uw}{\partial z}
= -\rho\frac{\partial\Phi}{\partial x}+2\Omega v\rho\,,
$$

$$
\frac{\partial \rho v}{\partial t}
+\frac{\partial \rho uv}{\partial x}
+\frac{\partial (\rho v^2+P)}{\partial y}
+\frac{\partial \rho vw}{\partial z}
= -\rho\frac{\partial\Phi}{\partial y}-2\Omega u\rho\,,
$$

$$
\frac{\partial \rho w}{\partial t}
+\frac{\partial \rho uw}{\partial x}
+\frac{\partial \rho vw}{\partial y}
+\frac{\partial (\rho w^2+P)}{\partial z}
= -\rho\dfrac{\partial\Phi}{\partial z}\,,
$$

$$
\frac{\partial \rho E}{\partial t}
+\frac{\partial \rho uh}{\partial x}
+\frac{\partial \rho vh}{\partial y}
+\frac{\partial \rho wh}{\partial z}
= -\rho u\frac{\partial\Phi}{\partial x}
-\rho v\frac{\partial\Phi}{\partial y}
-\rho w\frac{\partial\Phi}{\partial z}\,.
$$
Here, $\rho$ denotes density; $u$, $v$, and $w$ are the $x$,
$y$, and $z$ components of the velocity vector ${\bmath
v}=(u,v,w)$; $P$ is the pressure;
$E=\varepsilon+\slantfrac{1}{2}\cdot|{\bmath v}|^2$ is the
total specific energy;
$h=\varepsilon+P/\rho+\slantfrac{1}{2}\cdot|{\bmath v}|^2$ is the
total specific enthalpy; and $\Phi$ is the Roche potential:

$$
\Phi({\bmath r}) = -\frac{G M_1}{|{\bmath r} - {\bmath r}_1|}
-\frac{G M_2}{|{\bmath r} - {\bmath r}_2|}
-\slantfrac{1}{2}\Omega^2{\left({\bmath r}-{\bmath r}_c\right)}^2\,,
$$
$\Omega=2\pi /P_{orb}$ is binary's angular velocity; ${\bmath
r}_1$, ${\bmath r}_2$ is the radius-vectors of centers of mass of
system components; and ${\bmath r}_c$ is the radius-vector of
center of masses of a binary system.  The calculations were
carried out in the non-inertial Cartesian coordinate system
rotating together with the binary system.  Origin of coordinates
is located in the center of a mass-losing component, `$x$'-axis
is directed along the line connecting the centers of stars, from
the mass-losing component to the accretor, `$z$'-axis is
directed along the axis of rotation, and $y$'-axis is determined
so that we obtain a right-hand coordinate system. To close the
system of equations, we used the equation of state of ideal gas
$P=(\gamma-1)\rho\varepsilon$, where $\gamma$ as it usually is
the ratio of heat capacities. To mimic the system with radiating
losses, we accept in the model the value of adiabatic index
close to unit:  $\gamma = 1.01$, that corresponds to the case
close to the isothermal one [12,13].

\renewcommand{\thefigure}{1}
\begin{figure}[t]
\centerline{\hbox{\psfig{figure=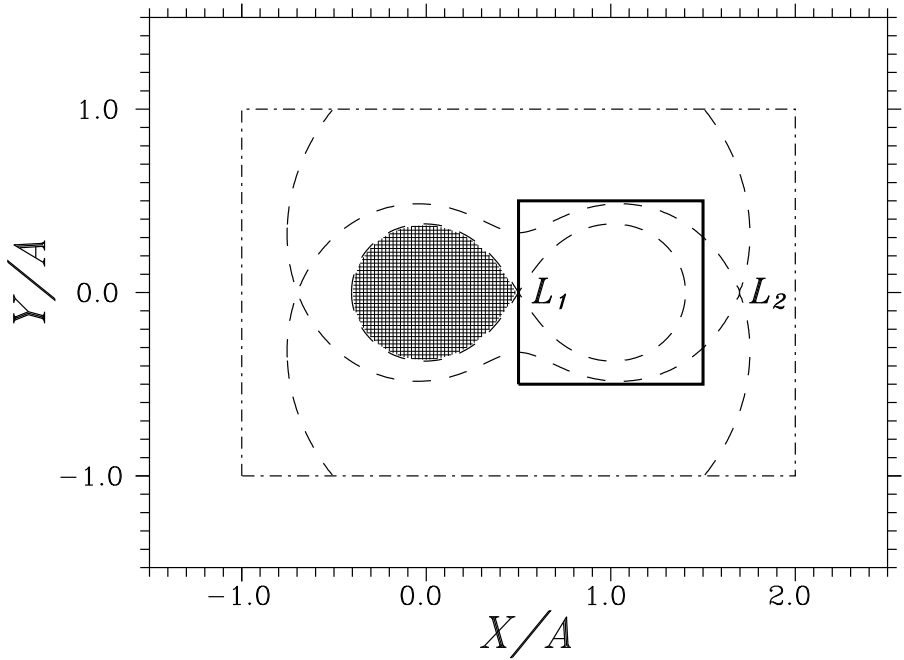,width=12cm}}}
\caption{\small
The calculation domains for the `complete' (dash-dotted line) and
`restricted' (bold line) problems. Shaded region is the donor.
The Roche equipotentials are shown by dashed lines.}
\end{figure}

\renewcommand{\thefigure}{2a}
\begin{figure}[t]
\centerline{\hbox{\psfig{figure=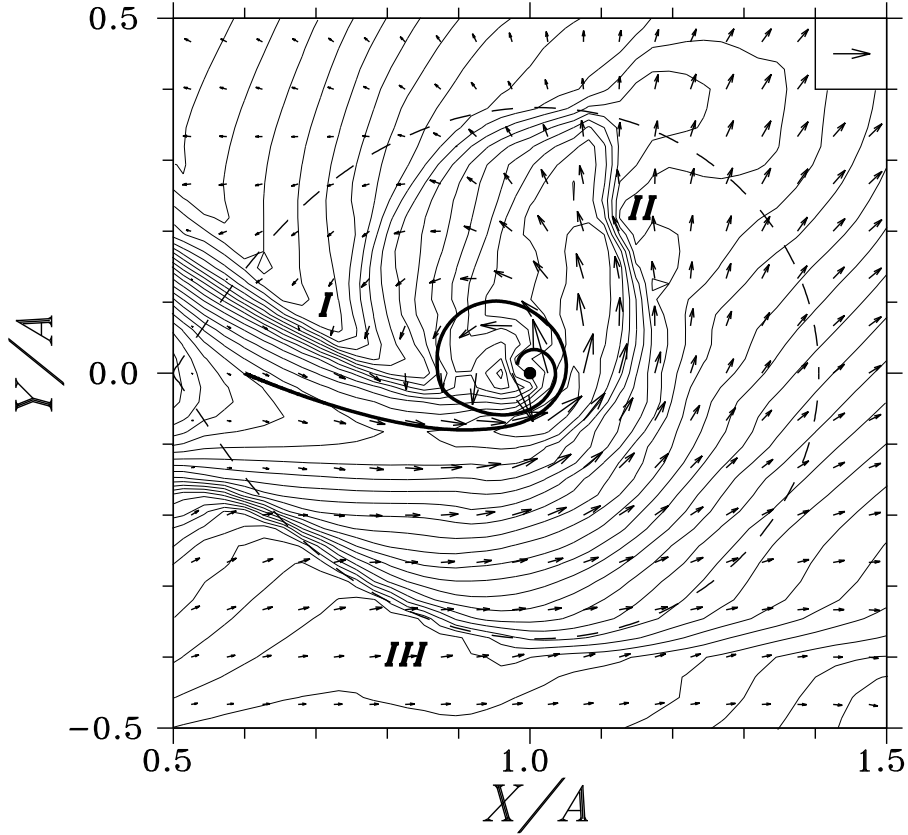,width=12cm}}}
\caption{\small
Isolines of density and velocity vectors in the equatorial
plane for the run in the `complete' region (see [6]). Filled
circle is the accretor, dashed lines are the Roche
equipotentials. The vector in the upper right corner corresponds
to dimensionless velocity $v=3$.  Bold line is the boundary
(`marginal') flowline, along which the matter falls in the
disk.}
\end{figure}

\renewcommand{\thefigure}{2b}
\begin{figure}[t]
\centerline{\hbox{\psfig{figure=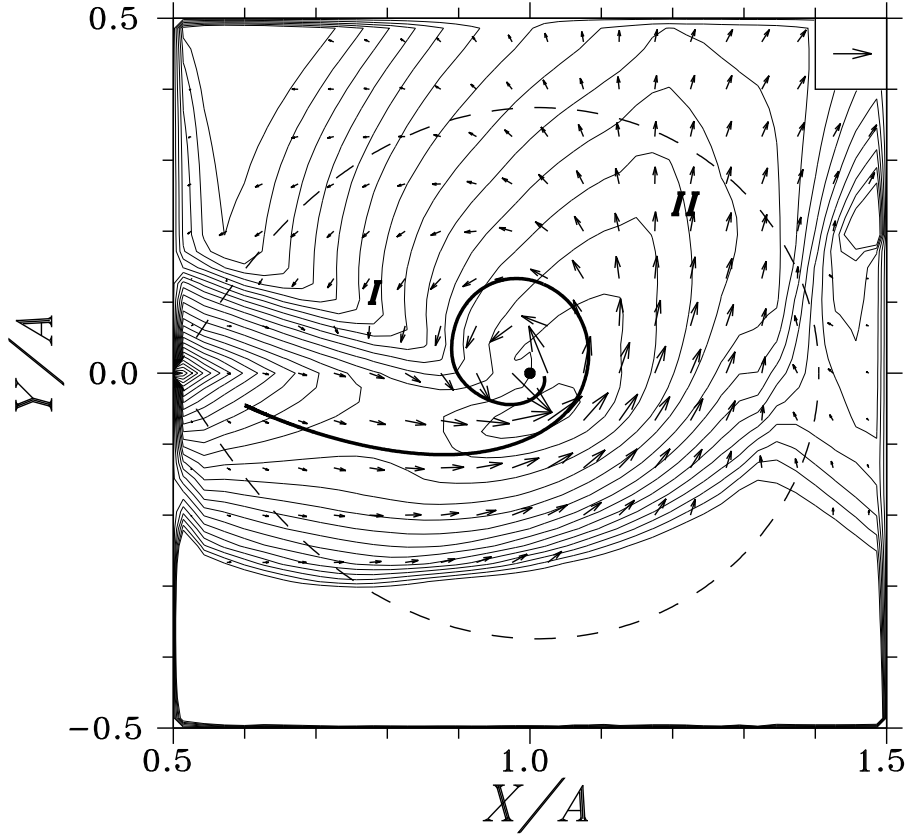,width=12cm}}}
\caption{\small
Isolines of density and velocity vectors in the equatorial plane
for the run in the `restricted' calculation domain on the same
grid as that shown on Fig.~2a. Filled circle is the accretor,
dashed lines are the Roche equipotentials. The vector in the
upper right corner corresponds to dimensionless velocity $v=3$.
Bold line is the boundary (`marginal') flowline, along which the
matter falls in the disk.}
\end{figure}

\renewcommand{\thefigure}{3}
\begin{figure}[t]
\centerline{\hbox{\psfig{figure=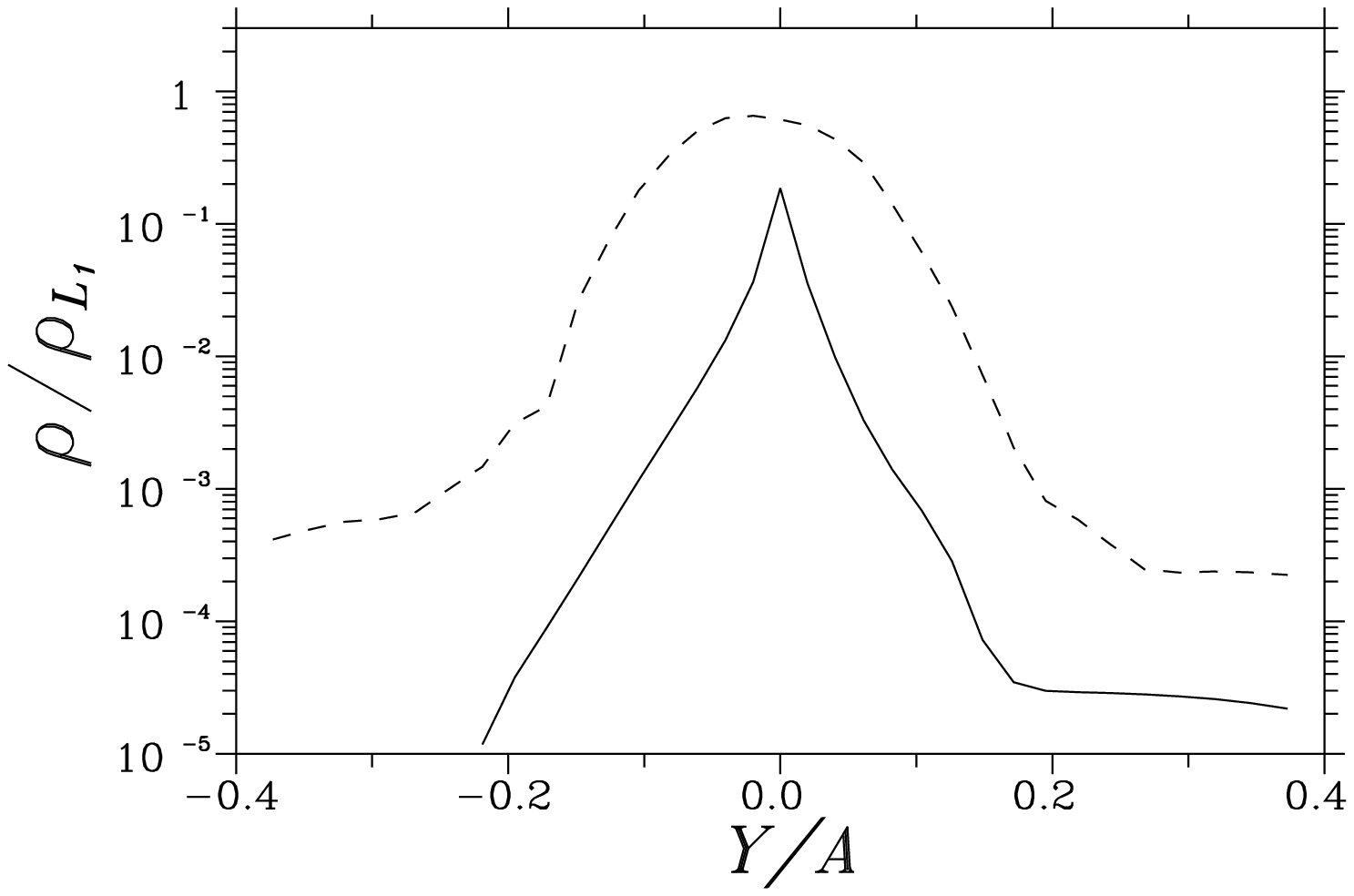,width=12cm}}}
\caption{
One-dimensional density profiles for
`complete' (dashed line) and `restricted' runs (solid line).
The density values are taken along the line in the
equatorial plane passing near the $L_1$ point parallel to
the `$y$'-axis ($x=\slantfrac{1}{2}+h$, $z=0$).}
\end{figure}

\renewcommand{\thefigure}{4}
\begin{figure}[t]
\centerline{\hbox{\psfig{figure=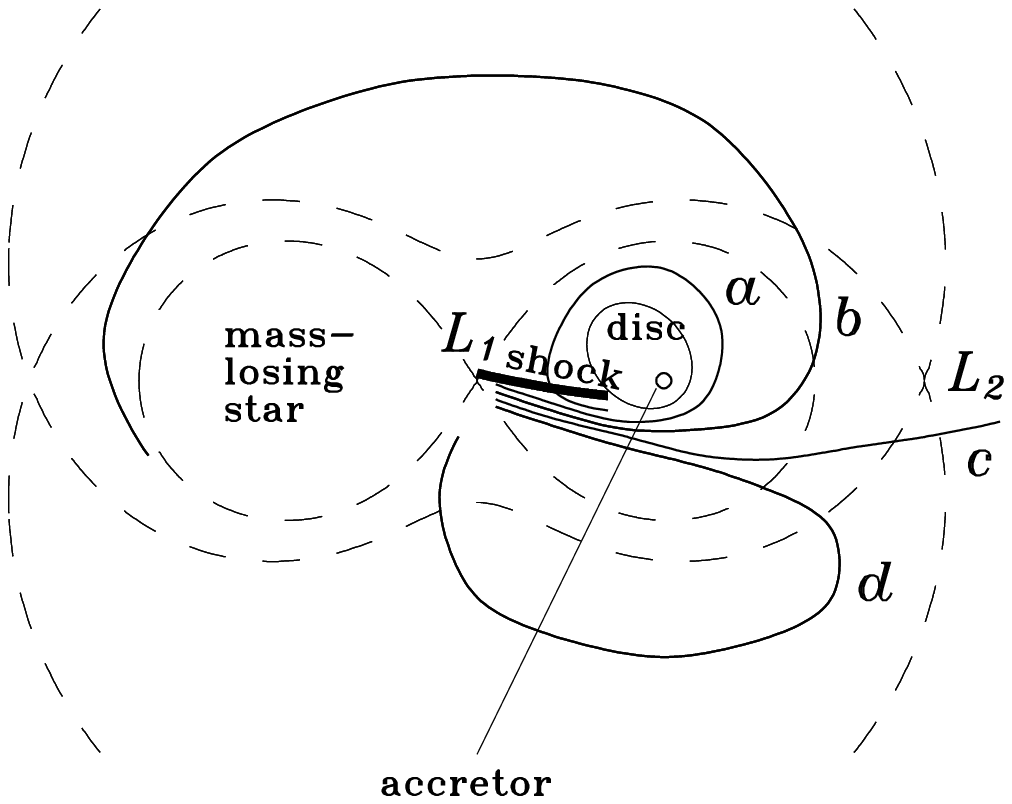,width=10cm}}}
\caption{\small
Schematic presentation of gas dynamical flow patterns in
semidetached binary systems. The Roche lobe (dashed lines), the
accretor, the Lagrangian points, and the quasi-elliptic
accretion disk are shown. Bold line is the shock wave formed as
a result of interactions between the circumbinary envelope gas
and the stream. The flowlines $a$, $b$, $c$, $d$ show the
directions of gaseous flows in the system. Flowlines $a$, $b$,
$d$ form the circumbinary envelope of the system, flowline $c$
is the gas flow leaving the system.}
\end{figure}

\renewcommand{\thefigure}{5a}
\begin{figure}[t]
\centerline{\hbox{\psfig{figure=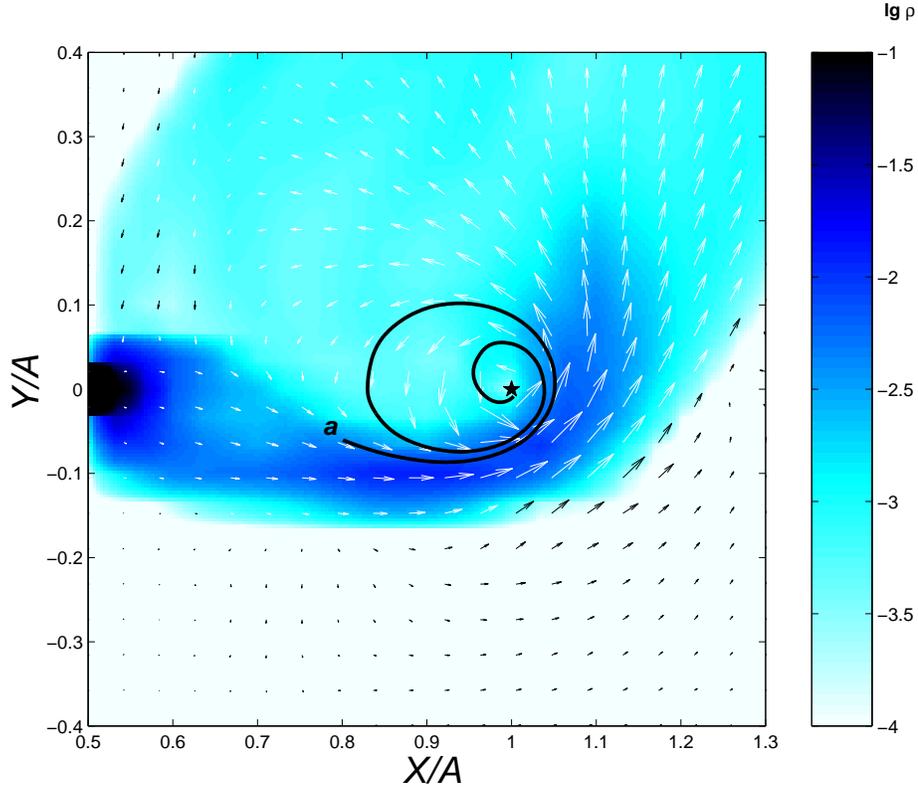,width=12cm}}}
\caption{\small
The distribution of density over the equatorial plane for
run `A' (grid $31\times31\times17$). Arrows are the
velocity vectors. Black asterisk is the accretor. Flowline $a$
bounding the accretion disk is also shown. Correspondence
between density logarithm and gradation of grey
color is shown on the scale.}
\end{figure}

\renewcommand{\thefigure}{5b}
\begin{figure}[t]
\centerline{\hbox{\psfig{figure=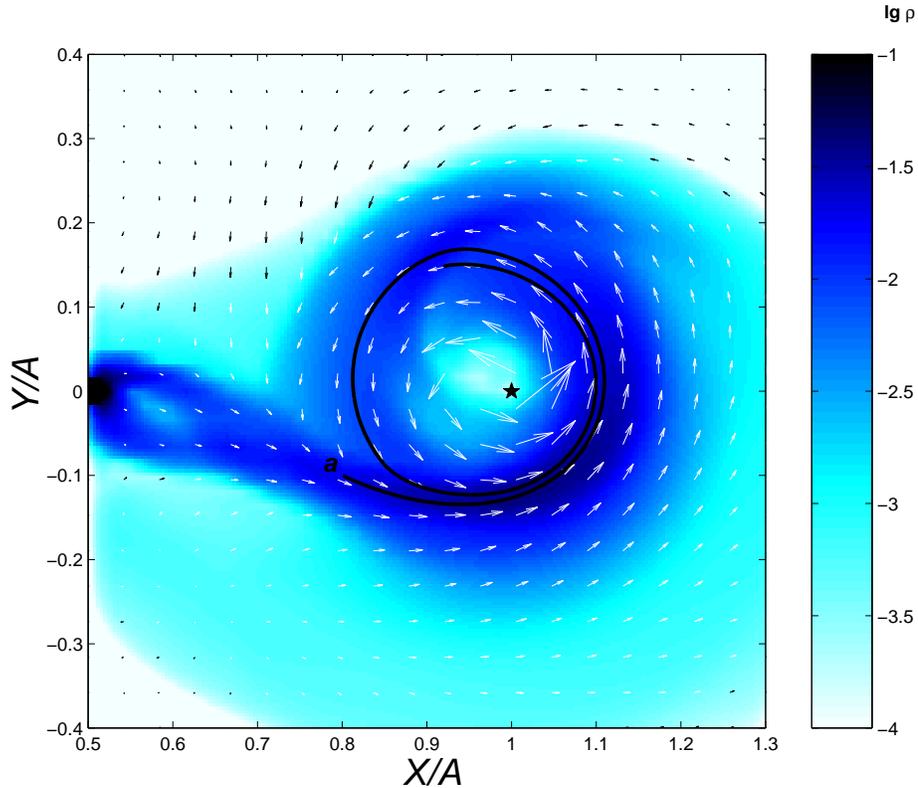,width=12cm}}}
\caption{\small
The same as in Fig.~5a for run `B' (grid
$61\times61\times17$).}
\end{figure}

\renewcommand{\thefigure}{5c}
\begin{figure}[t]
\centerline{\hbox{\psfig{figure=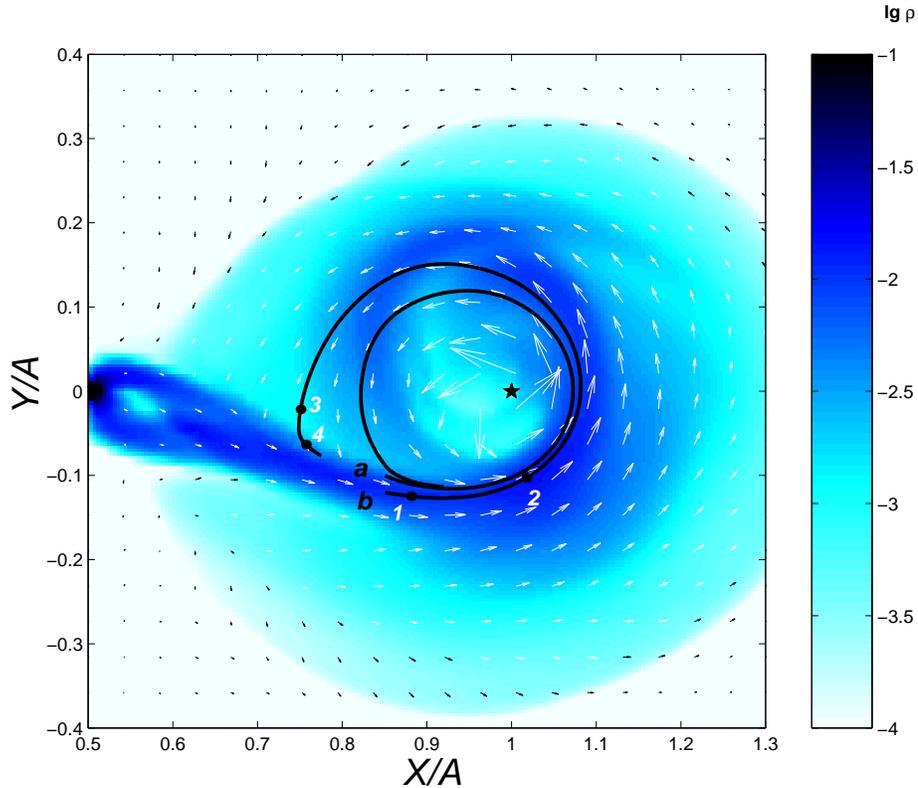,width=12cm}}}
\caption{\small
The same as in Fig.~5a for run `C' (grid
$91\times91\times25$). In addition to flowline $a$ bounding
the accretion disk, flowline $b$ passing through the shock
is also shown. Zone `1-2' of flowline $b$
corresponds to the place where gas stream from $L_1$
contacts with accretion disk, zone `3-4' corresponds to the
place of interaction between circumbinary gas and the stream
(shock wave $I$).}
\end{figure}

\renewcommand{\thefigure}{6}
\begin{figure}[p]
\centerline{\hbox{\psfig{figure=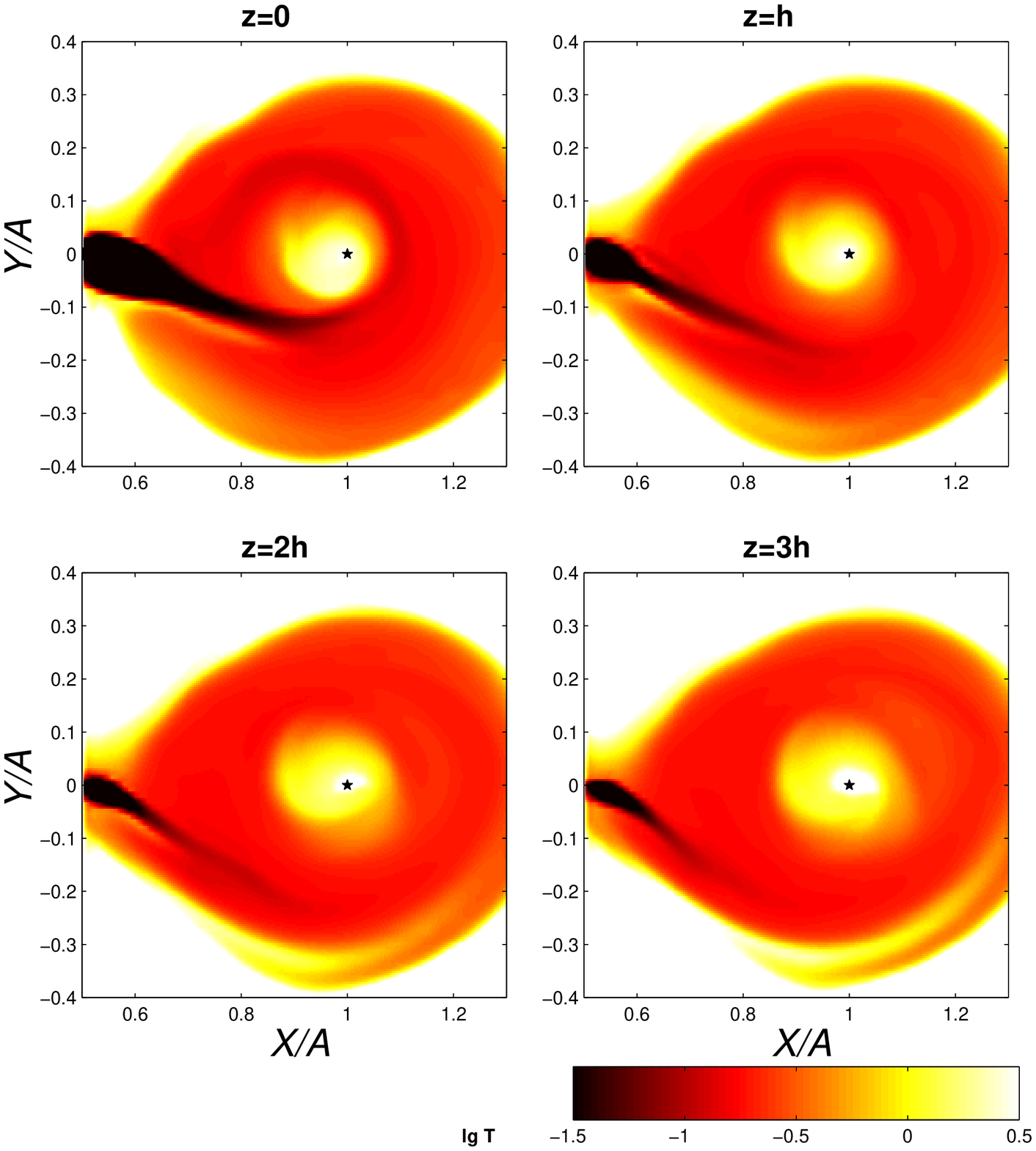,width=15cm}}}
\caption{\small
The distribution of dimensionless temperature over equatorial
plane $z=0$ and the parallel planes $z=h$, $z=2h$, $z=3h$ for
run `C' (grid $91\times91\times25$). Here $h$ is the height of
the gridcell that is approximately equal to $0.01A$. Black
asterisk is the accretor.}
\end{figure}

\renewcommand{\thefigure}{7}
\begin{figure}[t]
\centerline{\hbox{\psfig{figure=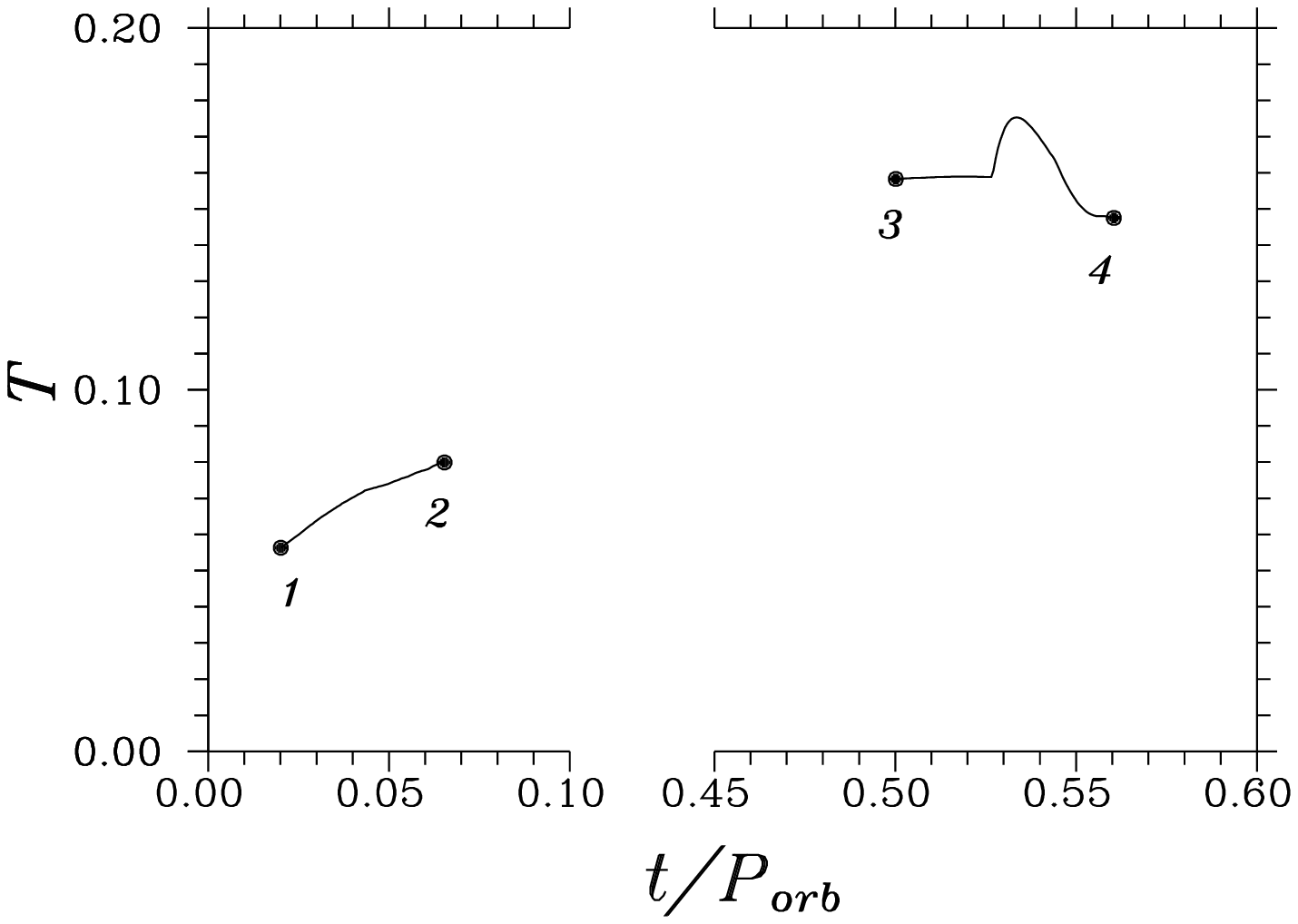,width=10cm}}}
\caption{\small
Dimensionless temperature along flowline $b$, ref. Fig.~5c.
Zone `1-2' corresponds to the place of contact of the stream and
the accretion disk, zone `3-4' corresponds to the shock wave
$I$.}
\end{figure}

\renewcommand{\thefigure}{8}
\begin{figure}[t]
\centerline{\hbox{\psfig{figure=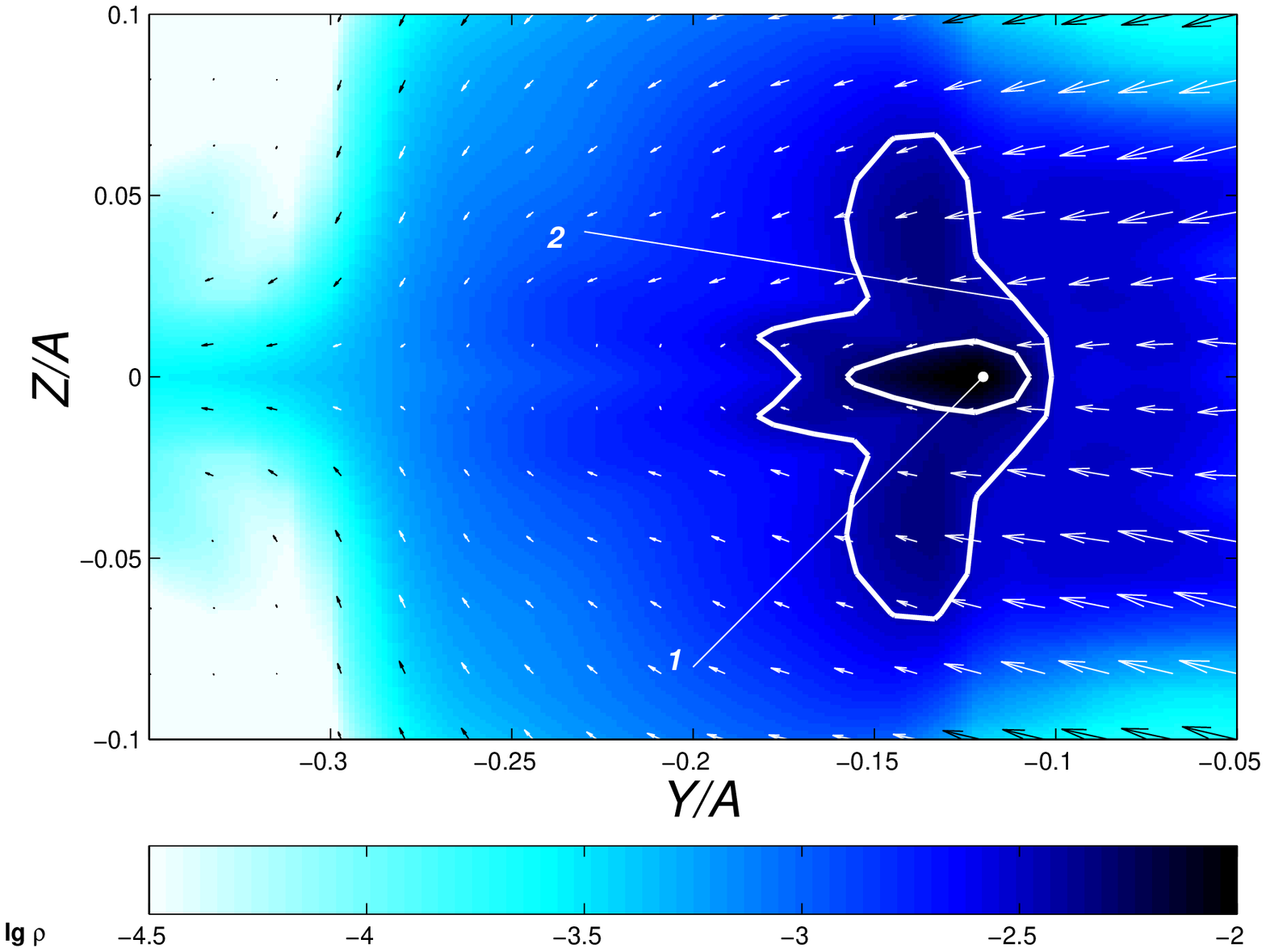,width=12cm}}}
\caption{\small
Distribution of density and velocity vectors in `$yz$'-plane
passing through point $(\slantfrac{9}{10},0,0)$, for
run `C' (grid $91\times91\times25$). Marker 1 designates
the central part of the stream limited by isoline
$\lg\rho=-2.3$. Marker 2 on isoline $\lg\rho=-2.5$ shows the
front of shock wave $I$.}
\end{figure}

\renewcommand{\thefigure}{9}
\begin{figure}[t]
\centerline{\hbox{\psfig{figure=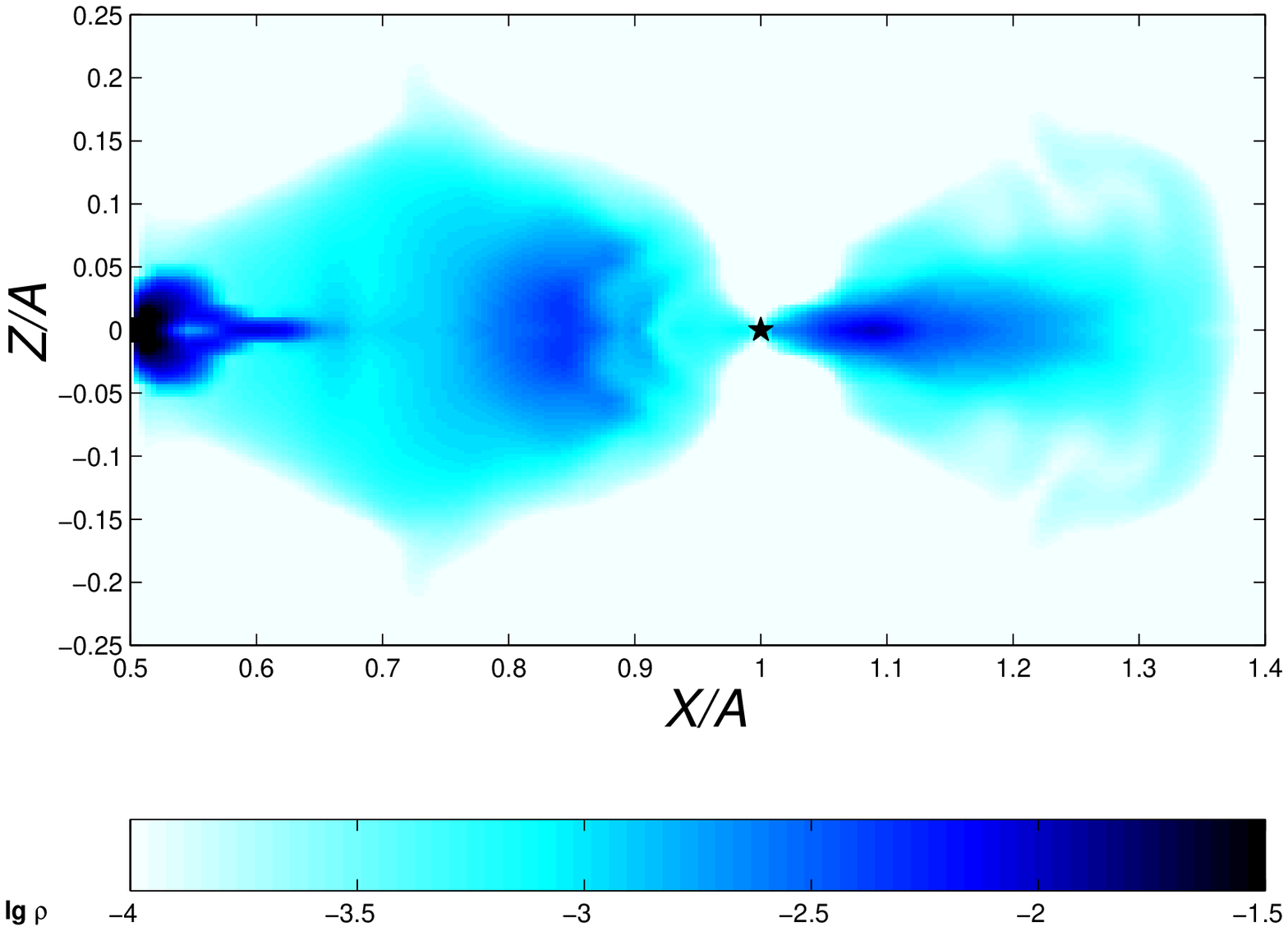,width=12cm}}}
\caption{\small
The distribution of density and velocities vectors in frontal
plane (`$xz$'-plane) for run `C' (grid $91\times91\times25$).
Black asterisk is the accretor.}
\end{figure}

\renewcommand{\thefigure}{10a}
\begin{figure}[t]
\centerline{\hbox{\psfig{figure=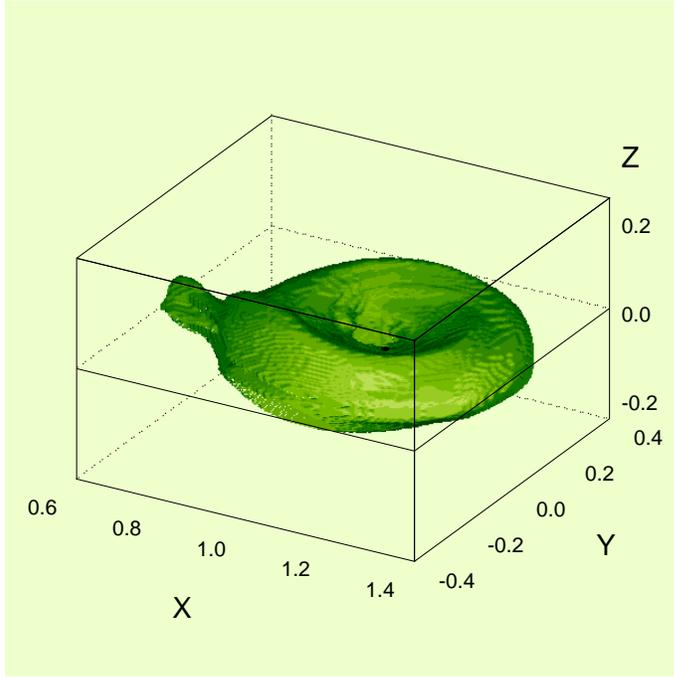,width=9cm}}}
\caption{\small
3D image of density iso-surface at level $\rho=10^{-3}$. Black
circle is the accretor.
}
\end{figure}

\renewcommand{\thefigure}{10b}
\begin{figure}[p]
\centerline{\hbox{\psfig{figure=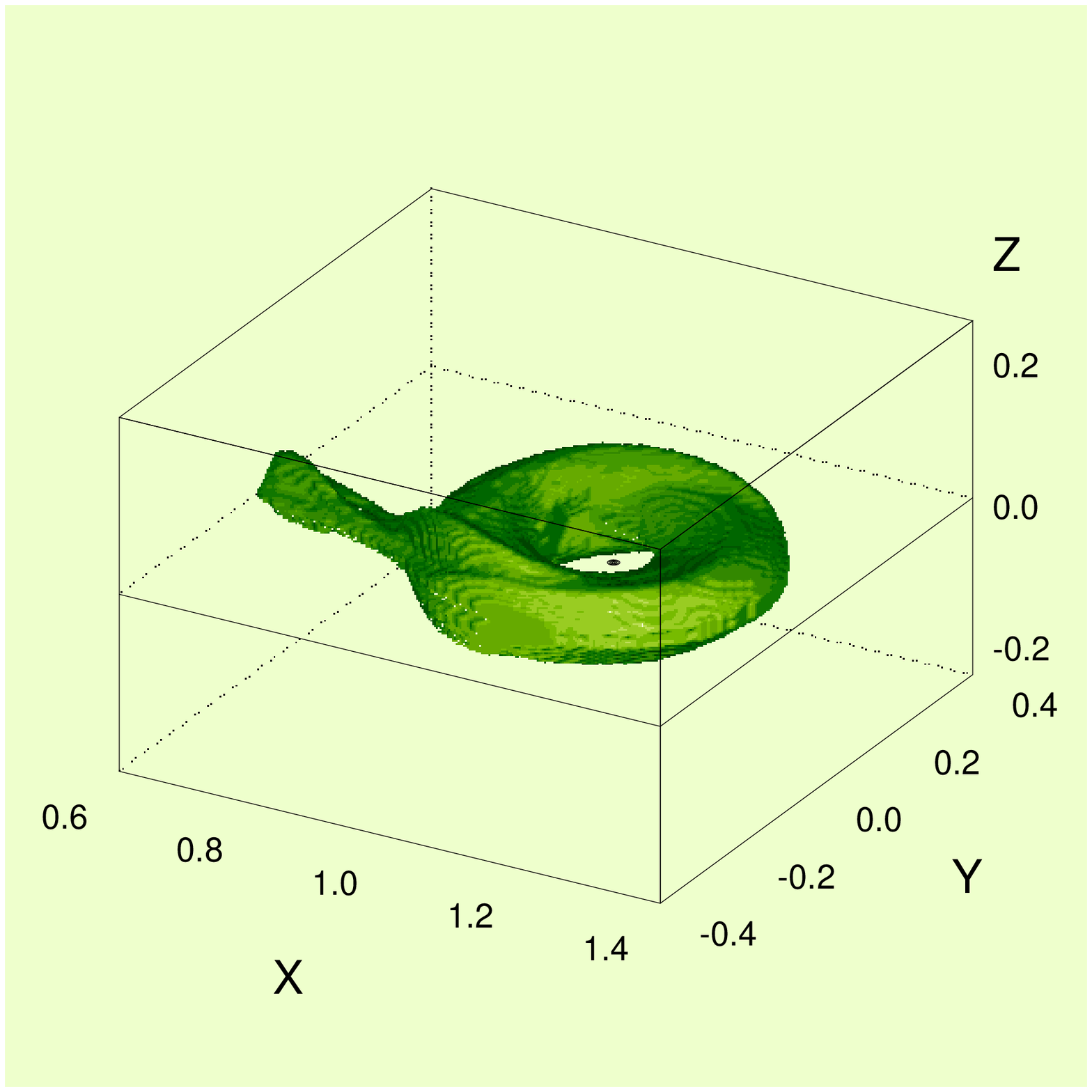,width=9cm}}}
\caption{\small
The same as in Figure 10a for density $\rho=3\cdot10^{-3}$.}
\end{figure}

\renewcommand{\thefigure}{10c}
\begin{figure}[p]
\vspace*{10mm}
\centerline{\hbox{\psfig{figure=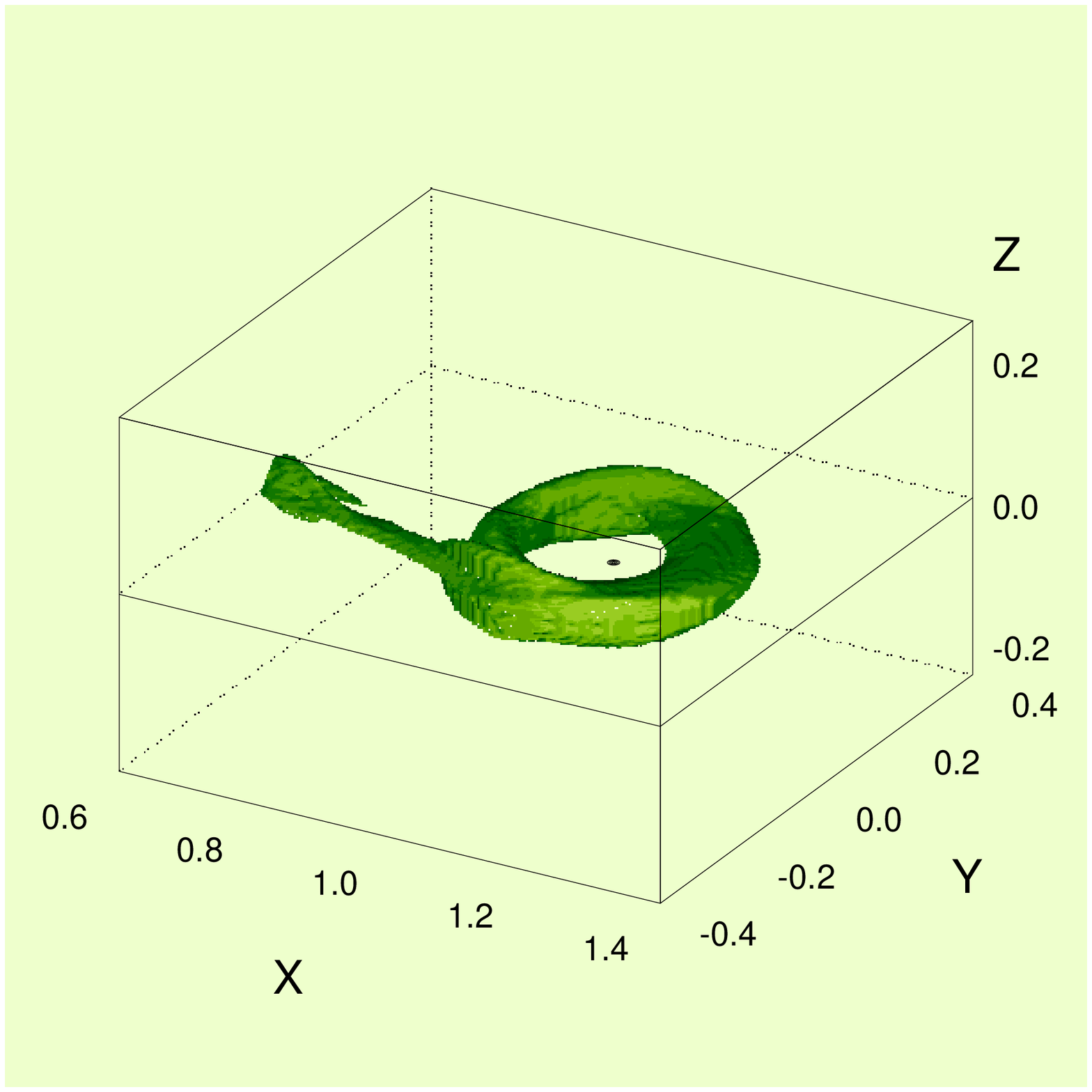,width=9cm}}}
\caption{\small
The same as in Figure 10a for density $\rho=6\cdot10^{-3}$.
}
\end{figure}

To obtain numerical solution of the system of equations we used
the Roe--Osher TVD scheme of a high approximation order [14,15]
with Einfeldt modification [16]. The original system of
equations was written in a dimensionless form. To do this, the
spatial variables were normalized to the distance between the
components $A$, the time variables were normalized to the
reciprocal angular velocity of the system ${\Omega}^{-1}$, and
the density was normalized to its value\footnote{Because the
system of equations can be scaled to density and pressure, the
density scale was chosen simply for the sake of convenience.} in
the inner Lagrangian point $L_1$.

As it was noted above we decreased numerical viscosity by
changing of the gridcell size. Unfortunately, the limited power
of the computer used did not allow us to simulate the flow over
a large domain (as in the calculations in [1--6] where the size
of computational domain was 3 times larger than the binary
system separation) on a fine grid.  Therefore, similarly to
works [17--20], the gas flow was simulated over a restricted
domain that was set as parallelepipedon
$[\slantfrac{1}{2}A\ldots
\slantfrac{3}{2}A]\times[-\slantfrac{1}{2}A\ldots
\slantfrac{1}{2}A]\times[0\ldots \slantfrac{1}{4}A]$
(calculations were conducted only in the top half-space). A
sphere with a radius of \slantfrac{1}{100} representing the
accretor was cut out of the calculation domain. By way of
illustration, the parameters of the calculation domain in the
equatorial plane are shown in Fig.~1 (solid line); the dashed
curves are the Roche equipotentials, and the dashed-dotted lines
show the boundary of the calculation domain adopted in  [1--6].

As the initial conditions we used rarefied gas with the
following parameters $\rho_0=10^{-5}$, $p_0=10^{-4}/\gamma$,
${\bmath v}_0=0$. The boundary conditions were determined as
follows: we preset the conditions in a fictitious gridpoint
corresponding to the inner Lagrangian point:  $\rho(L_1)=1$,
$p(L_1)=10^{-2}/\gamma$, $u(L_1)=10^{-2}$, $v(L_1)=w(L_1)=0$;
the velocity of sound in this gridpoint was equal to
$c(L_1)=10^{-1}$. In fictitious points located inside of the
accretor and also in the fictitious points on the outer border
except $L_1$ point all the parameters were taken equal to the
initial (`background') values. The final (numerical) boundary
conditions were determined by solving the Riemann problem
between the gas parameters in fictitious gridpoints and the gas
parameters in the closest gridpoint.

\section{RESULTS}

In [1--6] we studied the morphology of flow over a complete
computational domain that is few times larger than binary
separation $A$ and includes both system components (see
Fig.~1). In these calculations the circumbinary envelope was
found to play an important role in formation of the
flow structure.  The use of a restricted calculated domain in
numerical model precludes the account of the circumbinary
envelope and, hence, the `restricted' solution can appreciably
differ from the `complete' one. Let us consider the effect of
the restricted calculated domain on the solution. We compare the
structure of flow from the `complete' model [6] and the
calculation obtained at the same grid in the restricted domain.
Figures~2a and~2b depict isolines of density and the
velocity vectors in the equatorial plane of the system for the
`complete' and the `restricted' calculations, respectively.
Figures~2a and~2b also represent the boundary (`marginal')
flowline of the stream, along which the matter falls directly in
the disk. Comparison of the results proves that the morphology
of flows in the region near the disk in the `restricted' problem
reflects the basic features of flow patterns of the `complete'
solution. In particular, an accretion disk forming in the system
has approximately the same linear size and shock waves $I$ and
$II$ appear as a result of interaction of gas of the
circumbinary envelope with the stream from $L_1$ (note that
hereinafter the position of shock waves is determined by
condensing of density isolines and the presence of mass flow
through the surface). Meanwhile, in the `restricted' problem the
effect of the circumbinary envelope is taken into account not
quite correctly. In particular, matter flows along the Roche
lobe of a donor-star that result in the stripping of the
atmosphere of a donor-star and essential increase in the rate of
mass transfer in the system are not taken into account (see
[2]).  This fact causes essential change in the parameters of
the stream of the `restricted' problem in comparison to the
`complete' one (see Fig.~3 where one-dimensional distributions of
density along a line parallel to `$y$'-axis and lying in
equatorial plane are shown for `complete' and the `restricted'
calculations).  In particular, in the `restricted' problem the
stream flowing out from $L_1$ tends to extend to the
characteristic value $\epsilon=c(L_1)/A\Omega$ [21], and the
density decreases along the cross-section of the stream by
exponential law, while in the `complete' problem the density of
flowing matter is considerably determined by the stripping
effect of the atmosphere of the donor-star. Moreover, the
`restricted' problem does not take account of the flow of the
circumbinary gas returning in the system under the impact of
Coriolis force (from the side of the stream which is opposite to
the orbital motion). As a results, shock wave $III$ disappears
in the `restricted' problem but exists in the `complete' one.
Summarizing the results of comparison the `complete'
(the basic features of the
obtained flow structure for `complete' formulation are
summarized on the schematic diagram  -- Fig.~4) and the
`restricted' problems, we can state that:

\begin{itemize}

\item In the `restricted' problem the effect of the circumbinary
envelope on the solution is considered not quite correctly. As a
results, it causes, from one hand, qualitative changes of the
solution: a part of flows of the circumbinary gas returning in
the system is absent in the `restricted' solution (flowlines
$d$ and $b$ in Fig.~4), and, hence, shock wave $III$ disappears;
on the other hand, quantitative changes:  the parameters of
stream change because the stripping effect is not taken into
account, and, in particular, the thickness of the stream tends
to decrease.

\item The morphology of gaseous flows in the vicinity of the
accretor in the `restricted' and the `complete' solutions is
similar at a qualitative level. The partial account of the
circumbinary envelope in the `restricted' problem (e.g.,
the matter revolving around the accretor and
interacting with a stream -- see flowline $a$ in Fig.~4) allows
one to obtain the solution similar to the `complete' one. In
particular, both in the `complete' problem (see [1--6]) and in
the `restricted' solution the flow deflecting under the impact
of gas of the circumbinary envelope approach to the disk at a
tangent line and does not cause shock disturbances of the edge
of disk (`hot spot'). In both solutions the regions of
superfluous energy release are located along the edge of stream
facing towards the orbital motion where the interaction of the
circumbinary envelope with the stream causes extended shock wave
$I$ to form (see Figs~2,4).

\end{itemize}

The comparison of the `restricted' and the  `complete' problems
discussed above proves that stream--disk interaction can also be
considered with some clauses in the `restricted' problem.
Therefore, we can conduct the fine grid calculations and, hence,
to get the solutions at lower viscosity.  Accordingly, we can
study the impact of viscosity value on the flow structure in the
vicinity of the accretor and consider the problem of `hot spot'
appearance at low viscosity.

To study the effect of viscosity we consider the results of
three calculations with the various
spatial resolution:  $31\times31\times17$, $61\times61\times17$,
and $91\times91\times25$ (hereinafter, `A', `B' and `C' runs,
respectively). In all the calculations the grid was taken
uniform. In terms of $\alpha$-disk, the numerical viscosity of
`A', `B' and `C' runs corresponded to $\alpha\sim 0.08\div0.1$,
$0.04\div0.06$, and $0.01\div0.02$.

Comparison of the results of `A', `B' and `C' runs among each
other and with the results of study [20], where grid
$200\times200\times50$ was used within the framework of the same
problem setup allows one to study the impact of viscosity on the
solution.  Figures~5a,~5b and~5c show the fields of
density and the velocity vectors in the equatorial  plane of the
system for `A', `B' and `C' variants and also flowline $a$
bounding the accretion disk. Numerical analysis of the obtained
solutions proves the shock-free interaction of the stream and
the disk in all variants of calculations. The morphology of the
stream-disk system is uniform and, hence, the `hot spot' does
not form. This fact is obvious from Fig.~6 showing the
distributions of dimensionless temperature in `$xy$'-plane for
run `C' (at the minimal value of viscosity) for four values of
`$z$'-coordinate:  $z=0$ -- the equatorial plane, $z=h$, $z=2h$,
and $z=3h$, where the height of the gridcell $h\sim0.01A$. The
dimensionless temperature in point $L_1$ is equal to $10^{-2}$.
To obtain the dimensional temperature one should multiply the
dimensionless value by $GM/AR$, where $G$ is the gravitational
constant, $M$ is the total mass of the system, $A$ is the system
separation, $R$ is the gas constant.  Analysis of the
temperature fields shown in Fig.~6 proves the absence of the
energy release zone in the place of contact of a stream and a
disk, i.e. the absence of `hot spot' at all values of
`$z$'-coordinate. Consideration of the change in the gas
parameters along the flowlines supports this conclusion. In
particular, Fig.~7 depicts the change of dimensionless
temperature along flowline $b$ (Fig.~5c), from which one may
conclude that `hot spot' is absent in the place of contact of a
stream and a disk (zone `1-2') and that the energy release
region locates in the place of forming shock wave $I$ (zone
`3-4' in Figs.~5c and~7)

Comparative analysis of the obtained solutions proves (see,
e.g., Fig.~5a,b,c) that increasing of spatial resolution and,
hence, reduction of numerical viscosity causes reduction of
diffusion spreading of the stream.  In the runs with a thinner
stream (`B' and `C' runs, and the calculation in work [20]) gas
of the circumbinary envelope tends to flow around the stream
from above and from below.  An illustration of this fact can be
seen in Fig.~8 where the density field and the velocity
vectors in `$yz$'-plane passing through point
$(\slantfrac{9}{10},0,0)$ are shown for `C' run.  The change of
density (the increase in flow thickness) in the region where gas
of the circumbinary envelope interacts with the stream is also
obvious in Fig.~9 represented the density field in frontal
plane (`$xz$'-plane) for run `C'.  The presence of the gas
flowing around a stream may give the
impression that shock interaction appears between the stream and
the disk (see, for example, the three-dimensional image of
iso-density surface in Fig.~10, and also the work [20]), though
analysis of the solution unambiguously proves the smooth nature
of this part of the flow.

In our studies [1--6] we structured the flow patterns as follows:

\begin{itemize}

\item disk is the matter of stream that immediately captured by
the accretor and then is accreted;

\item circumbinary envelope is all the rest matter. The part of
this matter revolves around the accretor, interacts with the
stream and then can be accreted.

\end{itemize}

Thus, we divided the matter of stream by a physical attribute:
if gas leaves the system or then interacts with the initial
stream, we consider this matter as not belonged to the disk. In
the solutions with small viscosity there was additional part of
the envelope: the matter that flows around a stream from above
and from below. Analysis of the flowlines proves that this part
of the envelope does not belong to the disk. The large part of
this matter leaves the system or, rotating around the accretor
gradually comes nearer to the equatorial  plane and then it does
collide with the flow. In addition to the definitions of
works [1--6], it is worthwhile introducing an additional term:
the `circumbinary halo' describing the matter that: i) revolves
around the accretor (it is gravitationally captured); ii) does
not belong to the disk; iii) interacts with the stream (collides
or flows around from above and from below); iv) then, after
interaction, it can be either involved in accretion or leaves
the system.

The change in numerical viscosity should inevitably result in
the change of accretion rate.  Analysis of `A', `B' and `C'
runs, as it can be expected, proves that in the calculations
with smaller viscosity the rate of accretion decreases.
Unfortunately, the detailed study of this question requires huge
computing time that essentially exceeds the power of the
computer used. The problem is that in the calculations with high
viscosity that were presented in [1--6] the viscous
time scale

$$
\tau^{visc}
=\frac{R_{disk}^2}{\nu}
=\frac{R_{disk}^2}{\alpha c_s H}\,,
$$
where $c_s$ is the velocity of sound, $H$ is the characteristic
thickness of a disk,
exceeded the gas dynamical time scale for the disk

$$
\tau^{gas-dyn}
=\frac{R_{disk}}{c_s}\,,
$$
only slightly, that allowed one to reach the steady-state
solution at the times of several orbital periods. At decreased
viscosity the steady-state solution can hardly be obtained
because restricted power of the computer used does not allow one
to conduct the calculations for the periods of time larger than
$\tau^{visc}$.  Analysis of `C' variant, i.e. the study
of the mass of the disk and halo as functions of time, proves
that the solution was not a steady-state even at the times
larger than 15 orbital periods.

\section*{CONCLUSIONS}

Obtained results of 3D numerical simulations of mass transfer in
semidetached binaries demonstrate the absence of the `hot spot'
in self-consistent solution. This conclusion was made
when considering systems with large value of viscosity (see
[1--6]). Presented calculations confirm that there is no shock
interaction between the stream from $L_1$ and the forming
accretion  disk (`hot spot') at any value of viscosity.

Analysis of obtained solutions shows that in the systems with
various values of viscosity the stream from $L_1$ is deflected
under the impact of gas of the circumbinary envelope and
approaches to the disk at a tangent line, so does not cause
shock disturbances of the edge of disk (`hot spot'). The regions
of superfluous energy release are located along the edge of
stream facing towards the orbital motion. So we can conclude
about the qualitative similarity of the flow structure for
various values of viscosity. At the same time runs with low
value of viscosity (when diffusion is weak and thickness of the
stream is small) reveal an important role of the gas of the
circumbinary envelope that flows around the stream from above
and from below.  We have introduced an additional term to
describe this feature of the flow structure: the `circumbinary
halo'.

\section*{ACKNOWLEDGMENTS}
This work was supported by the Russian Foundation for Basic
Research (grant 99-02-17619) and by grant of President of Russia
(99-15-96022).

\section*{REFERENCES}

\small

\begin{enumerate}

\item Bisikalo, D.V., Boyarchuk, A.A., Kuznetsov, O.A. \&
Chechetkin, V.M. 1997, Astron. Zh., 74, 880 (Astron. Reports,
41, 786; astro-ph/9802004)

\item Bisikalo, D.V., Boyarchuk, A.A., Kuznetsov, O.A. \&
Chechetkin, V.M. 1997, Astron. Zh., 74, 889 (Astron. Reports,
41, 794; astro-ph/9802039)

\item Bisikalo, D.V., Boyarchuk, A.A., Kuznetsov, O.A.,
Khruzina, T.S., Cherepa\-shchuk, A.M. \&  Chechetkin, V.M. 1998,
Astron. Zh., 75, 40 (Astron. Reports, 42, 33; astro-ph/9802134)

\item Bisikalo, D.V., Boyarchuk, A.A., Chechetkin, V.M.,
Kuznetsov, O.A. \& Molteni, D. 1998, Monthly Notices Roy.
Astron. Soc., 300, 39 (astro-ph/9805261)

\item Bisikalo, D.V., Boyarchuk, A.A., Kuznetsov, O.A. \&
Chechetkin, V.M. 1998, Astron. Zh., 75, 706 (Astron. Reports,
42, 621; astro-ph/9806013)

\item Bisikalo, D.V., Boyarchuk, A.A., Chechetkin, V.M.,
Kuznetsov, O.A. \& Molteni, D. 1999, Astron. Zh. (in press;
astro-ph/9907084)

\item Meyer, F. \& Meyer-Hofmeister, E. 1981, Astron.
Astrophys., 104, L10

\item Cannizzo, J.K., Chen, W. \& Livio, M. 1995, Astrophys. J.,
454, 880

\item Honma, F., Matsumoto, R. \& Kato, S. 1993, Astrophys. J.
Suppl., 210, 365

\item Szuszkievich, E. \& Miller, J.C. 1997, Monthly Notices
Roy. Astron. Soc., 287, 165

\item Cannizzo, J.K. 1993, Astrophys. J., 419, 318

\item Sawada, K., Matsuda, T. \& Hachisu, I. 1986, Monthly
Notices Roy. Astron. Soc., 219, 75

\item Bisikalo, D.V., Boyarchuk, A.A., Kuznetsov, O.A., Popov,
Yu.P. \& Che\-chetkin, V.M. 1995, Astron. Zh., 72, 367 (Astron.
Reports, 39, 325)

\item Roe, P.L. 1986, Ann. Rev. Fluid Mech., 18, 337

\item Osher, S. \& Chakravarthy, S. 1984, SIAM J. Numer. Anal.,
21, 955

\item Einfeldt, B. 1988, SIAM J. Numer. Anal., 25, 294

\item Molteni, D., Belvedere, G. \& Lanzafame, G. 1991, Monthly
Notices Roy. Astron. Soc., 249, 748

\item Lanzafame, G., Belvedere, G. \& Molteni, D. 1992, Monthly
Notices Roy. Astron. Soc., 258, 152

\item Armitage, P.J. \& Livio, M. 1996, Astrophys. J., 470, 1024

\item Makita, M., Miyawaki, K. \& Matsuda, T. 1999, Monthly
Notices Roy. Astron. Soc. (submitted; astro-ph/9809003)

\item Lubow, S.H. \& Shu, F.H. 1975, Astrophys. J., 198, 383

\end{enumerate}

\end{document}